\documentclass[aps,preprint]{revtex4}%
\usepackage{amsfonts}
\usepackage{amsmath}
\usepackage{amssymb}
\usepackage{graphicx}%
\providecommand{\U}[1]{\protect\rule{.1in}{.1in}}
\begin{document}
\title[ ]{Hydrogen Molecular Ion and Molecule in Classical Electrodynamics with
Classical Zero-Point Radiation}
\author{Timothy H. Boyer}
\affiliation{Department of Physics, City College of the City University of New York, New
York, New York 10031}

\begin{abstract}
The hydrogen molecular ion and the hydrogen molecule are treated in an
approximate calculation based on classical electrodynamics which includes
classical zero-point radiation. \ It is found within the classical theory that
a molecular ion is less-well bound than a hydrogen atom plus a distant proton.
\ The smaller binding energy is explained due to the repulsive nature of the
force between the proton and the atom when considering the most natural
resonant orbit of the electron. \ The approximate classical electromagnetic
calculation gives a binding energy of $2.1eV$ and a proton separation of
$0.916A^{o}$ for the hydrogen molecular ion. \ The hydrogen molecule is formed
by adding a single additional electron to the ion. \ The electrostatic
attraction of the electron to the ion gives a binding energy of $4.6eV$ and a
inter-proton separation of $0.6A^{o}$ for the hydrogen molecule in this
classical electromagnetic approximation.\ 

\end{abstract}
\email{tboyer@ccny.cuny.edu}
\maketitle

\section{Introduction}

Old quantum theory seemed to account for the nonrelativistic hydrogen atom,
even relativistic aspects, but the theory stumbled badly when attempting to
account for the hydrogen molecular ion and for the hydrogen molecule ground
state.\cite{Pais} \ Old quantum theory used classical particle trajectories
but fixed values for the action variables $J_{i}$ of an atomic system;
however, the theory contained no aspect involving randomness or resonance.
\ In the present article, we treat the hydrogen molecular ion and then the
hydrogen molecule using only \textit{classical} electrodynamics including
Lorentz-invariant classical zero-point radiation.\cite{Review} \ The classical
theory contains an element of randomness, which can be written in terms of
random phases for the zero-point radiation.\cite{Rice} \ Thus, the randomness
of the theory might seem analogous to the randomness of the positions and
velocities of gas molecules in a closed container. \ Just as for the situation
involving the randomness for the gas molecules, we are not able to predict the
outcome of a particular event, but we can give the average behavior of a group
of atoms.

There is a limitation in the present article. \ The full calculation would
involve a \textit{computer} \textit{simulation calculation} using random
zero-point radiation.\cite{CZ}\cite{HB} \ In the present article, we will find
only the equilibrium ground state orbit associated with complete energy
balance, where the energy radiated away by a charge particle is equal to the
average energy absorbed from the random classical zero-point radiation. The
analysis is more complicated than that for the hydrogen atom since the system
now involves three charges (in the case of $H_{2}^{+}$) or four charges (in
the case of $H_{2})$ which are all actually emitting and absorbing radiation.
\ We will make the standard approximation usually made for the hydrogen
molecular ion, regarding the two protons as so much more massive than the one
electron that we can imagine that they do not move compared to the electron
motions.\cite{GriffithsQ}\cite{ER} \ There is only one moving electron in
$H_{2}^{+}$ and only two simultaneously moving electrons for neutral $H_{2}$. \ 

\section{Connections to Old Quantum Theory}

\subsection{Resonance}

Old quantum theory assigned the values of the action variables as $J_{i}%
=n_{i}\hbar$ where $n_{i}$ is an integer. \ However, a \textit{classical
electromagnetic} treatment including classical electromagnetic zero-point
radiation\cite{B2026c} emphasizes that these values for the particle angular
momentum arise from \textit{resonance} between periodic orbital motion and
random classical zero-point radiation. \ The resonance occurs from repeated
small gains of energy from motion in a periodic classical orbit. \ These small
gains balance the energy lost by the charge due to radiation emission. \ This
need for repeated periodic motion leading to resonance was not appreciated in
old quantum theory. \ Thus, Pauli's doctoral thesis\cite{Pauli} considered the
hydrogen molecular ion $H_{2}^{+}$, but he did not restrict his view to simple
periodic motions, but rather considered very complicated classical motions.
\ The hydrogen molecular ion permits many orbits for fixed values of
$J_{r},J_{\phi},J_{z}$ which seem to be only conditionally periodic and so
would not correspond to resonance situations with classical zero-point
radiation. \ \ In the first quarter of the 20th century, the $H_{2}^{+}$ ion
was an important example convincing physicists of the failure of old quantum
theory. \ 

\subsection{Classical Electromagnetic Resonance}

It is easy to find the approximate \textit{equilibrium} situation for the
hydrogen molecular ion in classical electrodynamics with classical
electromagnetic zero-point radiation. \ The ion consists of two separate
protons at a fixed separation $R$ and one electron in classical
electromagnetic zero-point radiation. \ In equilibrium, the electron
experiences a Coulomb attraction to both the two stationary protons, and has a
horizontal distance $l=R/2$ from each of the protons. \ In order to give a
bound system, the electron must be between the protons, experiencing a Coulomb
attraction to both protons which cancels the electrostatic repulsion between
the protons. \ 

We might assume that the electron is stationary so that the problem becomes
electrostatic. \ However, if random classical zero-point radiation is present,
it will provide random forces tending to destabilize any attempt at an
\textit{electrostatic} solution. \ \ It seems natural to assume that, in the
energy equilibrium situation, the electron is in a resonant \textit{circular}
orbit, so that the whole situation has axial symmetry around the axis formed
by the line connecting the two protons. \ 

Since the electron is moving in a circle, it is accelerating and hence losing
energy through radiation emission. \ It must gain energy from the repeated
random impulses of the classical zero-point radiation. \ Since the classical
zero-point radiation has action variable $J_{rad}=\hbar$ no matter what the
frequency of the radiation $\omega,$ we expect the average action variable for
the electron in its ground state to be given by $J=\hbar$. \ And this
situation leads to an energy balance between power lost and average power
gained by the electron. \ 

We consider only the \textit{nonrelativistic} limit where the speed of light
in vacuum $c$ is taken as very large, and classical zero-point radiation
therefore becomes merely a random electromagnetic fluctuation in amplitude and
frequency. \ The average force (average in time) on the protons must vanish at
equilibrium. \ 

\subsection{Similarities to Old Quantum Theory}

Lorentz-invariant classical zero-point radiation has one single scale $\left[
U^{zp}\left(  \omega\right)  \right]  /\omega$ for each radiation normal mode
of frequency $\omega$, independent of frequency, so that $U^{zp}%
(\omega)=\left(  1/2\right)  \hbar\omega.$ \ Here $\hbar$ is the one scale
factor inherited by all the action variables from classical electromagnetic
zero-point radiation and is the same for all frequencies $\omega$. \ Thus, as
shown in previous articles,\cite{B} the (average) electron action variable
$J_{i}$ takes the values $J_{i}=n_{i}\hbar$ where $n_{i}$ is an integer. In
classical electromagnetic theory, the integer values arise because of resonant
gain of energy from classical zero-point radiation. \ The assignment
$J_{i}=n_{i}\hbar$ is the same as in old quantum theory, but the restriction
to periodic orbits allowing resonance is new. \ Thus the theory of classical
electrodynamics with classical zero-point radiation has similarities to old
quantum theory, but, in the assumption of randomness, is also quite different. \ 

\section{Outline of the Analysis}

\subsection{Formation of the Hydrogen Molecular Ion}

When the separation between the protons is large, the electron will be
attracted to one of the protons and form a hydrogen atom, while the other
proton remains far away. \ However, the distant proton will polarize the
hydrogen atom.\ The circular electron orbit can be expected to assume an
orientation approximately perpendicular to the line joining the two protons.
\ When the distant proton is at a much greater distance than the Bohr orbit of
the atom, there will be \textit{repulsion} between the atom and the proton
because the distance between the protons is smaller than the distance from the
distant proton to the electron in its Bohr orbit. \ We can imagine external
forces applied to the protons so as to slowly move them together despite the
repulsion. \ The electron orbit will have a place between the two protons such
that the component of force along the line joining the two protons is the same
in magnitude for each proton.

As the external proton comes closer, one becomes aware that there are two
different orbits which involve the same horizontal distance from the protons.
\ One electron orbit is the ever-smaller orbit followed by the electron which
is being pulled out from its near-by proton. \ This smaller orbit involves a
repulsion between the two protons and a smaller energy for the system of three
particles. \ The other \textit{possible} orbit for the electron has a very
large radius and is coming in from spatial infinity. \ This large orbit would
involve an attraction between the protons. \ As the protons come still closer,
the radius of the electron orbit becomes still smaller, as does the larger
possible electron orbit coming in from infinite radius. \ At some separation
between the protons, the smaller electron orbit reaches its smallest radius.
\ \ At this distance, due to the random impulses of the classical zero-point
radiation, the fast-moving electron will radiate away part of its energy and
move to the possible orbit of large radius where the total energy is lower and
the electron velocity is smaller.\ \ The hydrogen molecular ion has been
formed with the orbiting electron half-way between the protons. \ \ \ \ \ \ \ \ \ \ 

\subsection{Pulling the Hydrogen Molecular Ion Apart}

If one starts with the electron in a hydrogen molecular ion, one expects that
the electron will be in a circular orbit half-way between the two protons, in
an orbital plane which is perpendicular to the line joining the protons. \ Of
course, the random classical zero-point radiation will cause variations in
this picture, but the requirement of resonance means that the eccentricity is
expected to be small so as to give periodic motion.\ 

If one imagines equal and opposite forces on the two protons, these will tend
to pull the ion apart in a symmetrical fashion. \ Even in equilibrium, the
radius of the electron orbit is larger than the electron orbital radius of the
electron in hydrogen. \ Thus, as the protons are pulled apart, the radius of
the electron orbit is expected to become ever \textit{larger}. \ However, at
some point the fluctuations of the random classical zero-point radiation
should bring the electron closer to one proton than the other. \ At this
distance, we expect the electron to radiate away its additional energy and to
join the nearer proton as a hydrogen atom with the other proton some distance
away. \ 

\section{Calculation of the Ground State Energy Balance Orbit}

\subsection{Equations of Electron Motion}

In the hydrogen molecular ion, the electron located at $\mathbf{r}$ in the
$xy$-plane with radius $r$ from the axis of symmetry $\widehat{z}$ and at
angle $\phi=\omega t$ is pictured as following a circular orbit in time $t$
\begin{equation}
\mathbf{r}\left(  t\right)  =\widehat{x}r\cos\left(  \omega t\right)
+\widehat{y}r\sin\left(  \omega t\right)
\end{equation}
situated symmetrically between the two stationary protons located on the
$z$-axis at $\pm l$. \ The nonrelativistic Newtonian equation of motion for
the centripetal acceleration of the electron with speed $v=\omega r$ and
separation $l$ of the protons from the plane of the electron orbit is%
\begin{equation}
M\frac{v^{2}}{r}=2\frac{e^{2}}{\left(  r^{2}+l^{2}\right)  }\frac{r}%
{\sqrt{r^{2}+l^{2}}}=\frac{2e^{2}r}{\left(  r^{2}+l^{2}\right)  ^{3/2}}
\label{Mv2dr}%
\end{equation}
where the two protons provide the centripetal force, \ 

The repeated resonant motion of the electron in its circular orbit provides
the angular momentum $J$ with average value $\hbar$. \ Thus, the
nonrelativistic angular momentum $J$ of the electron in its orbit is along the
$z$-axis and is given by
\begin{equation}
J=\hbar=rMv\text{ \ \ or \ \ }v=\hbar/\left(  rM\right)  \label{JerMv}%
\end{equation}
Then we have from Eqs. (\ref{Mv2dr}) and (\ref{JerMv})%
\begin{equation}
\frac{M}{r}\left(  \frac{\hbar}{rM}\right)  ^{2}=\frac{2e^{2}r}{\left(
r^{2}+l^{2}\right)  ^{3/2}}%
\end{equation}
or simply%
\begin{equation}
\frac{\hbar^{2}}{Me^{2}}=r_{B}=\frac{2r^{4}}{\left(  r^{2}+l^{2}\right)
^{3/2}} \label{h2Me2}%
\end{equation}
where $r_{B}$ is the radius of the Bohr orbit in hydrogen,
\begin{equation}
r_{B}=\hbar^{2}/\left(  Me^{2}\right)  =0.529A^{o}.
\end{equation}

\subsection{Equilibrium for the Hydrogen Molecular Ion\ }

At steady state, the net force along the axis of symmetry for each proton
should vanish. \ This requirement gives for each proton%
\begin{equation}
0=\frac{-e^{2}}{\left(  r^{2}+l^{2}\right)  }\frac{l}{\sqrt{r^{2}+l^{2}}%
}+\frac{e^{2}}{\left(  2l\right)  ^{2}}%
\end{equation}
or equivalently%
\begin{equation}
4l^{3}=\left(  r^{2}+l^{2}\right)  ^{3/2}. \label{4R3}%
\end{equation}

It is easy to solve equations (\ref{h2Me2}) and (\ref{4R3}) numerically and to
find that the equilibrium situation involves an electron orbital radius
\begin{equation}
r=0.565A^{o}\text{ \ \ } \label{randl}%
\end{equation}
and a distance to each of the protons
\begin{equation}
l=0.458A^{o},
\end{equation}
giving a separation between the protons of
\begin{equation}
R=2l=0.916A^{o}. \label{2R}%
\end{equation}

The nonrelativistic classical electromagnetic energy of the hydrogen molecular
ion (ignoring any motion of the protons) involves both the kinetic energy of
the electron and the potential energy between all the electrostatic charges,
\begin{align}
\mathcal{E}  &  \mathcal{=}\frac{1}{2}Mv^{2}-2\frac{e^{2}}{\sqrt{r^{2}+l^{2}}%
}+\frac{e^{2}}{2l}=M\left(  \frac{\hbar}{rM}\right)  ^{2}-2\frac{e^{2}}%
{\sqrt{r^{2}+l^{2}}}+\frac{e^{2}}{2l}\nonumber\\
&  =e^{2}\left(  \frac{r_{B}}{2r^{2}}-\frac{2}{\sqrt{r^{2}+l^{2}}}+\frac
{1}{2l}\right) \nonumber\\
&  =e^{2}\left(  \frac{0.529}{2\left(  0.565\right)  ^{2}}-\frac{2}%
{\sqrt{\left(  0.565\right)  ^{2}+\left(  0.458\right)  ^{2}}}+\frac
{1}{2\left(  0.458\right)  }\right) \nonumber\\
&  =-\frac{e^{2}}{1.21A^{o}}=-e^{2}\left(  \frac{0.828}{A^{o}}\right)  .
\label{Eion}%
\end{align}
On the other hand, the classical electromagnetic energy of the Bohr atom in
its ground state is
\begin{equation}
\mathcal{E}_{Bohr}=-\frac{Me^{4}}{2\left(  \hbar\right)  ^{2}}=-\frac{Me^{4}%
}{2\left(  \hbar\right)  ^{2}}\left[  \left(  \frac{\hbar^{2}}{Me^{2}}\right)
\left(  \frac{1}{r_{B}}\right)  \right]  =-\frac{e^{2}}{2r_{B}}=-\frac{e^{2}%
}{1.058A^{o}}=-e^{2}\frac{0.945}{A^{o}}. \label{EBohr}%
\end{equation}
Thus we see that, according to classical electromagnetic theory, the
nonrelativistic energy of the hydrogen molecular ion is \textit{higher (less
deep)} than than the energy of a hydrogen atom and a proton, according to
classical electromagnetic theory with classical zero-point radiation. \ This
situation does not seem to be suggested in the standard textbooks of modern
physics. \ It is contrary to the diagram on page 420 of Eisberg and Resnick
which shows that hydrogen molecular ion as having a still lower energy than
the Bohr energy. \ But it does seem to be the result of classical physics with
classical zero-point radiation. \ 

\subsection{Simplifying the Electron Equations of Motion}

We are interested in comparing the two different situations, one situation
where the electron orbit is much closer to one proton than to the other, and a
second situation where the electron orbit exists midway between the protons.
\ The centripetal acceleration of the electron is \
\begin{equation}
M\frac{v^{2}}{r}=\frac{e^{2}r}{\left(  r^{2}+l^{2}\right)  ^{3/2}}+\frac
{e^{2}r}{\left(  r^{2}+L^{2}\right)  ^{3/2}}, \label{Mv2r}%
\end{equation}
while the component of force on the electron along the $z$-axis due to the
protons is%
\begin{equation}
0=\frac{e^{2}l}{\left(  r^{2}+l^{2}\right)  ^{3/2}}-\frac{e^{2}L}{\left(
r^{2}+L^{2}\right)  ^{3/2}}, \label{horiz}%
\end{equation}
and the magnitude of the $z$-component of force on the protons is%
\begin{equation}
\frac{e^{2}l}{\left(  r^{2}+l^{2}\right)  ^{3/2}}-\frac{e^{2}}{\left(
l+L\right)  ^{2}}=\frac{e^{2}L}{\left(  r^{2}+L^{2}\right)  ^{3/2}}%
-\frac{e^{2}}{\left(  l+L\right)  ^{2}}%
\end{equation}
where the separation between the protons is $R=l+L$. \ 

If we are searching for the situation of radiation energy balance between
emission and gain from zero-point radiation, we will assume in the
nonrelativistic approximation that $J=\hbar=rMv.$ Then equation (\ref{Mv2r})
becomes
\begin{equation}
\frac{M}{r}\left(  \frac{\hbar^{2}}{Mr}\right)  ^{2}=\frac{\hbar^{2}}{Me^{2}%
}\frac{e^{2}}{r^{3}}=\frac{0.529e^{2}}{r^{3}}=\frac{e^{2}r}{\left(
r^{2}+l^{2}\right)  ^{3/2}}+\frac{e^{2}r}{\left(  r^{2}+L^{2}\right)  ^{3/2}}%
\end{equation}
The factors of $e^{2}$ cancel on both sides of Eqs. (\ref{Mv2r})and
(\ref{horiz}) giving%
\begin{equation}
0.529=\frac{r^{4}}{\left(  r^{2}+l^{2}\right)  ^{3/2}}+\frac{r^{4}}{\left(
r^{2}+L^{2}\right)  ^{3/2}}. \label{0529}%
\end{equation}
and
\begin{equation}
\frac{l}{\left(  r^{2}+l^{2}\right)  ^{3/2}}=\frac{L}{\left(  r^{2}%
+L^{2}\right)  ^{3/2}}. \label{landL}%
\end{equation}

\section{Examples for the Hydrogen Molecular Ion}

\subsection{Distant Proton}

If we consider the situation where the external forces have moved the protons
to a separation where the distant proton has a horizontal separation from the
$xy$-plane $L=4A^{o},$ while the electron separated from the nearer proton is
$l/\left(  r^{2}+l^{2}\right)  ^{3/2}$, then we find the solutions of Eqs.
(\ref{0529}) and (\ref{landL}) are for $L=4A^{o},$ $r=0.528A^{o}$,
$l=0.00897A^{o}$. \ This solution gives an electrostatic repulsion of the
protons since
\begin{equation}
\frac{F_{p}}{e^{2}}=\frac{l}{\left(  r^{2}+l^{2}\right)  ^{3/2}}-\frac
{1}{\left(  l+L\right)  ^{2}}=-\frac{0.00131}{\left(  A^{o}\right)  ^{2}}.
\end{equation}
The energy of the situation is
\begin{equation}
\frac{U}{e^{2}}=-\frac{0.943}{A^{o}},
\end{equation}
which is slightly smaller that the energy of the nonrelativistic Bohr atom in
Eq. (\ref{EBohr}). \ The speed of the electron in its circular orbit is
\[
v\left(  r\right)  =v(0.528)=\left(  e^{2}/\hbar\right)  \left(
r_{B}/r\right)  \approxeq0.0073c.
\]

\subsection{Near Proton}

When the distance between the protons is shortened still more so that the
distant proton is separated by $L=1A^{o}$ from the plane of the electron
orbit, the solution of Eqs. (\ref{0529}) and (\ref{landL}) are $r=0.509A^{o},$
$l=0.0983A^{o},$ and the electrostatic repulsion of the protons has increased to%

\begin{equation}
\frac{F_{p}}{e^{2}}=-\frac{0.122}{\left(  A^{o}\right)  ^{2}},
\end{equation}
and the energy of the situation has decreased to
\begin{equation}
\frac{U}{e^{2}}=-\frac{0.889}{A^{o}}.
\end{equation}
The electron speed has increased to%
\begin{equation}
v\left(  0.509A^{o}\right)  =\left(  c/137\right)  \left(  0.529/0.509\right)
=0.00759c.
\end{equation}

\subsection{Transition}

The change may occur at $L=0.450A^{o}$ when the radius of the electron orbit
becomes $r=0.488A^{o}$ and the horizontal distance to the electron orbital
plane is $l=0.260A^{o},$ and the electrostatic repulsion of the protons has
increased to
\begin{equation}
\frac{F_{p}}{e^{2}}=-\frac{0.444}{\left(  A^{o}\right)  ^{2}}%
\end{equation}
and the energy has decreased to
\begin{equation}
\frac{U}{e^{2}}=-\frac{0.796}{A^{o}}.
\end{equation}
while the electron speed is
\begin{equation}
v\left(  0.488A^{o}\right)  =\left(  c/137\right)  \left(  0.529/0.488\right)
=0.00791c.
\end{equation}
It is in this region where the transition to equal spacing of the electron
orbit between the protons may occur. \ 

\subsection{Energy of the Molecular Ion}

We notice that the situation when $l=0.260A^{o},$ $L=0.450A^{o},$ and
$r=0.488A^{o}$ has a smaller negative energy $U=-e^{2}0.796/A^{o}$ than the
nonrelativistic hydrogen atom $U_{Bohr}=-e^{2}0.945/A^{o}$ by about
$0.945-0.796=0.149$ or $\Delta U=e^{2}0.149/A^{o}=2.1eV.$ \ This increase in
energy arises because of the repulsive nature of the electrostatic force
between the two protons. \ Thus the hydrogen molecular ion is less well bound
as compared to the hydrogen atom. \ The hydrogen molecular ion must overcome a
\textit{barrier} in order to relax to a still lower energy involving
nonrelativistic hydrogen atom and a distant additional proton. \ 

In contrast, the relaxation from the situation in the last example (where
$r=0.488A^{o},l=0.260A^{o},L=0.450A^{o}$) would involve the emission of
radiation from the small-radius-high-speed electron situation ($v(0.488A^{o}%
)=0.00791c$) over to the situation imagined in Eq. (\ref{randl}) involving a
large-radius-low-speed electron ($v(0.565A^{o})=0.00683c$). \ The
nonrelativistic angular momentum $J=\hbar=vMr$ is unchanged. \ However, the
energy has been decreased from $U=-e^{2}0.796/A^{o}$ over to $U=-e^{2}%
0.830/A^{o}$.

\section{Hydrogen Molecule}

\subsection{Formation of the Hydrogen Molecule from the Molecular Ion}

If two electrons are present, the natural positions, according to classical
physics, are on opposite sides of a circular orbit so as to get as far away
from each other as possible. \ On the other hand, we expect that the protons
would be pulled together by the presence of two attracting electrons. \ 

For the hydrogen molecule, the situation seems much easier if we already have
a hydrogen molecular ion. \ If a second electron is allowed to come near the
hydrogen molecular ion, it will be immediately attracted by the net plus
charge on the ion. \ In order to get as far way from the other electron as
possible while coming as close to the two protons as possible, we expect that
the second electron will rotate in the same circular electron orbit as the
first electron but on the opposite side of the circular orbit from the first
electron. \ Our only concern is the separation $l$ from the two protons and
the radius of the orbit of the two electrons. \ At equilibrium, we expect that
the net force on each proton vanishes and that the radius $r$ of the orbit of
the electrons gives the necessary balance between the centripetal force
provided by the protons and the centripetal acceleration of each electron.
\ The two quantities $r$ and $l$ are connected.

\subsection{Equations of Motion}

The balance between the forces tangential to the line joining the two protons
is immediate because a single distance $l$ is used. \ The centripetal motion
is given by Newton's second law,%
\begin{equation}
0.529=\frac{2r^{4}}{\left(  r^{2}+l^{2}\right)  ^{3/2}}-\frac{r}{4}.
\end{equation}
The energy of the nonrelativistic hydrogen molecule is%
\begin{equation}
\frac{U_{H_{2}}}{e^{2}}=2\left(  \frac{1}{2}\frac{(0.529)}{r^{2}}-\frac
{2}{\sqrt{r^{2}+l^{2}}}\right)  +\frac{1}{2l}+\frac{1}{2r}.
\end{equation}
The electromagnetic force on the left proton in the symmetric case is%
\begin{equation}
\frac{F_{H_{2}}}{e^{2}}=2\frac{l}{\left(  r^{2}+l^{2}\right)  ^{3/2}}-\frac
{1}{\left(  2l\right)  ^{2}}.
\end{equation}

\subsection{Equilibrium Situation}

The equilibrium situation involves a separation between the protons of
$R=2l=0.59A^{o}$ so that $l=0.295A^{o}$ while the radius of the orbit is
$r=0.510A^{o}$. \ Then the net force on each of the protons is very small
\begin{equation}
\frac{F_{H_{2}}}{e^{2}}=2\frac{0.295}{\left(  \left(  0.510\right)
^{2}+\left(  0.295\right)  ^{2}\right)  ^{3/2}}-\frac{1}{\left(  2\left(
0.295\right)  \right)  ^{2}}=\frac{0.014}{\left(  A^{o}\right)  ^{2}},
\end{equation}
while the energy is%
\begin{equation}
\frac{U_{H_{2}}}{e^{2}}=2\left(  \frac{1}{2}\frac{(0.529)}{\left(
0.510\right)  ^{2}}-\frac{2}{\sqrt{\left(  0.510\right)  ^{2}+\left(
0.295\right)  ^{2}}}\right)  +\frac{1}{2\left(  0.295\right)  }+\frac
{1}{2\left(  0.510\right)  }=-\frac{2.08}{A^{o}}.
\end{equation}
In order to convert to more familiar units, we write the energy as%
\begin{equation}
2.08\frac{(13.6)}{0.945}=29.93eV\text{ \ and \ \ }29.23-2\left(  13.6\right)
=2.73eV.
\end{equation}
This value is simply the naive value from the energy of the hydrogen molecule.
\ However, we should add on the binding energy of the hydrogen molecular ion
which is $2.1eV$ giving a total binding energy of $4.8eV.$ \ \ 

\section{Comparison with Accepted Values}

For the hydrogen molecular ion, the accepted binding energy is $2.7eV$ with a
separation between the two protons of $1.1A^{o}.$\cite{GriffithsQ}\cite{ER}
\ Our values are $2.1eV$ and $0.916A^{o}$. \ \ The discrepancies are about
22\% and 18\%. \ Such agreement for an approximate calculation might be
regarded as not unreasonable; a full calculation would involve a computer
simulation. \ Our values should be compared with those given by Griffiths'
quantum text\cite{GriffithsQ} for an approximate calculation, giving a binding
energy of $1.8eV$ at a separation of the protons of $1.3A^{o},$ corresponding
to discrepancies of $33\%$ and $18\%$ respectively.

For the hydrogen molecule, the accepted binding energy is about $4.5eV$ at a
separation between the two protons of $0.7A^{o}$.\cite{ER} \ Our values are
$4.8eV$ and $0.6A^{o}$, corresponding to discrepancies of\ $7\%$ and $14\%$
respectively. \ \ 

\section{Summary}

In this analysis of the hydrogen molecular ion and of molecular hydrogen, we
have used classical electrodynamics including classical electromagnetic
zero-point radiation in an approximate calculation. \ A full calculation would
involve a computer simulation with a model for random classical zero-point
radiation. \ Such simulations have been given in the past for the Coulomb
potential by Cole and Zou,\cite{CZ} and also for the harmonic oscillator by
Huang and Batelaan\cite{HB} \ He we assume that the component of angular
momentum along the line joining the two protons is a constant given by $\hbar$
since this is indeed the case in the simulations and since $\hbar$ is the only
constant which will fit the units of the action variable $J$, corresponding to
the angular momentum.

Our classical electromagnetic analysis suggests that the binding of the
hydrogen molecular ion is less deep than the binding of the electron in the
hydrogen atom. \ This difference is accounted for by the classical
electromagnetic repulsion of the proton at large distances when the electron
orbit is treated as a circular orbit pulled out from its near-by proton.
\ Within the classical electromagnetic picture given here, the hydrogen
molecule consists of two protons with two point electrons on opposite sides of
a circular orbit midway between the protons in a plane perpendicular to the
line joining the two protons. \ The values of the binding energy and the
equilibrium separation between the protons are within about 20\% of the
accepted experimental values. \ 

\section{Data Statement}

All the data associated with this article are contained within the article

August 15, 2026 \ \ \ \ \ \ \ \ H2+Ion4.tex

\end{document}